\documentclass[aps,superscriptaddress,twocolumn]{revtex4}
\usepackage{amsfonts,amssymb,amsmath}            
\usepackage[]{graphics,graphicx,epsfig}            
\usepackage{amsthm}
\def\identity{\leavevmode\hbox{\small1\kern-3.8pt\normalsize1}}

\newcommand{\ket}[1]{\left | #1 \right\rangle}
\newcommand{\bra}[1]{\left \langle #1 \right |}
\newcommand{\half}{\mbox{$\textstyle \frac{1}{2}$}}
\newcommand{\smallfrac}[2][1]{\mbox{$\textstyle \frac{#1}{#2}$}}
\newcommand{\Tr}{\text{Tr}}

\newcommand{\proj}[1]{\ket{#1}\bra{#1}}
\renewcommand{\epsilon}{\varepsilon}
\begin{document}

\title{Optimal Cloning and Singlet Monogamy}
\date{\today}

\author{Alastair \surname{Kay}}
\affiliation{Max-Planck-Institut f\"ur Quantenoptik, Hans-Kopfermann-Str.\ 1,
D-85748 Garching, Germany}
\affiliation{Centre for Quantum Computation,
             DAMTP,
             Centre for Mathematical Sciences,
             University of Cambridge,
             Wilberforce Road,
             Cambridge CB3 0WA, UK}
\affiliation{Centre for Quantum Technologies, National University of Singapore, 3 Science Drive 2, Singapore 117543}
\author{Dagomir Kaszlikowski}
\affiliation{Centre for Quantum Technologies, National University of Singapore, 3 Science Drive 2, Singapore 117543}
\affiliation{Department of Physics, National University of
Singapore, 2 Science Drive 3, Singapore 117542}
\author{Ravishankar Ramanathan}
\affiliation{Centre for Quantum Technologies, National University of Singapore, 3 Science Drive 2, Singapore 117543}
\begin{abstract}
The inability to produce two perfect copies of an unknown state is inherently linked with the inability to produce maximal entanglement between multiple spins. Despite this, there is no quantitative link between how much entanglement can be generated between spins (known as monogamy), and how well an unknown state can be cloned. This situation is remedied by giving a set of sufficient conditions such that the optimal Completely Positive map can be implemented as a teleportation operation into a standard, reference, state. The case of arbitrary $1\rightarrow N$ asymmetric cloning of $d$-dimensional spins can then be solved exactly, yielding the concept of `singlet monogamy'. The utility of this relation is demonstrated by calculating properties of Heisenberg systems, and contrasting them with the results from standard monogamy arguments.
\end{abstract}

\maketitle

{\em Introduction:} The inability to produce two perfect copies of an unknown state \cite{WZ82a} is an intrinsically quantum phenomenon whose profound implications have permeated a vast range of fields. For instance, the founding principle of quantum cryptographic schemes is that an eavesdropper cannot obtain a copy of any shared data without disturbing it in a detectable way, and this is guaranteed by the laws of physics rather than assumptions on the difficulty of computation. As these schemes generalize beyond key distribution to a variety of multiparty scenarios, it will become crucial to understand just what the limits on these disturbances are. Another fundamental concept of quantum mechanics, that of entanglement, has rather parallel properties. In particular, there is the notion of a maximally entangled state between two spins, and it is impossible for a specific spin to be maximally entangled with two other spins simultaneously. This concept of monogamy of entanglement has the potential to achieve a similar level of permeation, conceivably having an impact on fields as diverse as superconductivity \cite{dag}, but has so far failed to do so because, of the many different ways of measuring entanglement, a strict inequality relation has only been proven for the tangle \cite{tangle1,tangle}, and this particular measure is not a naturally applicable quantity in other branches of physics. Nevertheless, this inequality has proved useful for bounding ground state energies of some condensed matter systems.

Heuristically, the link between cloning and monogamy can be seen by considering a protocol whereby 3 spins are entangled. One follows a teleportation protocol with an unknown state, targeting spin 1. Copies of the unknown state appear on the other two spins, and the quality of the copies depends on how much entanglement was in the original state. Thus, if a particular quality of cloning is impossible, a certain degree of entanglement must be impossible. The inability to concisely express the conditions for optimal cloning of a single input spin onto $N$ copies ($1\rightarrow N$ cloning) in anything other than the fully symmetric case \cite{BH96a,GM97a,bruss-1998-57,BEM98a,werner1,werner2}, where all clones have the same output quality, has proved a major stumbling block to confirming any such expectations. The study of asymmetric cloning \cite{Cer00a,iblisdir-2005-72,iblisdir-2005,fiurasek-2005-5,rezakhani-2003} has largely been limited to special cases. An excellent review of the cloning problem can be found in \cite{rmp}.

In this paper, we demonstrate a sufficient condition on a CP map that allows one to determine its optimal implementation -- one can write down a matrix for which the maximum eigenstate is the optimal state to implement telemapping (the process where an input state is teleported into a pre-prepared state, and the outcome is the desired map), and the corresponding eigenvalue is the realized fidelity. $1\rightarrow N$ cloning is an interesting example which we solve in full generality. This allows the formulation of a monogamy-like relation, which we refer to as `singlet monogamy', for the singlet fractions
$$
p_{0,n}=\max_{U,V}\bra{B_0}U\otimes V\rho_{0,n}U^\dagger\otimes V^\dagger\ket{B_0}
$$
of the reduced states $\rho_{0,n}$ of a many-body state $\ket{\Psi}$, where $U$ and $V$ are arbitrary $d$-dimensional unitary rotations and $\ket{B_0}=\sum_{i=0}^{d-1}\ket{ii}/\sqrt{d}$.  Its utility is demonstrated by contrasting with the use of the original monogamy relation for qubits,
\begin{equation}
\sum_{n=1}^N\tau(\rho_{0,n})\leq\tau(\rho_{0,1\ldots N}), \label{eqn:tangle}
\end{equation}
in the calculation of ground state and thermal state properties, where $\tau$ is the tangle of the qubits \cite{tangle1,tangle}. That such a relation potentially offers significant advantages was observed in \cite{dag}.

{\em Sufficient Conditions for Optimal CP Maps:} In \cite{fiurasek-2001-64}, Fiur\'{a}\v{s}ek developed a general framework for determining an upper bound to the fidelity, $F$, of the conversion $\ket{\psi_{in}}\rightarrow\ket{\psi_{out}^{ideal}}$, where $\Lambda_{out}(\psi_{in})$ measures how well any output state $\rho_{out}^{actual}$ achieves the desired transformation due to a quantum map by averaging 
$$
\Tr\left(\sqrt{\sqrt{\rho_{out}^{actual}}\Lambda_{out}(\psi_{in})\sqrt{\rho_{out}^{actual}}}\right)^2
$$
over an arbitrary input distribution. If $\ket{\psi_{in}}$ is defined on a Hilbert space of dimension $d$, then $F\leq d\lambda$, where $\lambda$ is the maximum eigenvalue of the matrix
$$
R=\int d\psi_{in}\left(\proj{\psi_{in}}^T\otimes\Lambda_{out}(\psi_{in})\right),
$$
and $^T$ denotes the transpose. Telemapping, the generalization of telecloning, is the implementation of a suitable map by teleporting the state $\ket{\psi_{in}}$ into the input port(s) of a fixed state $\ket{\Psi}$ such that each of the output spins can be held by a separate party, each of whom is restricted to performing operations locally. We show that, under certain readily verifiable assumptions regarding the properties of correction operations, the bound of $F=d\lambda$ can be achieved, and the state $\ket{\Psi}$ required is the one that satisfies $R\ket{\Psi}=\lambda\ket{\Psi}$.

{\em Optimal Telemapping:} We are interested in using the $d^2$ Bell states of $d$ dimensional systems, $\ket{B_i}=(U_i\otimes\identity)\ket{B_0}$, to teleport a state $\ket{\psi_{in}}$ into the input port of a telemapping state $\ket{\Psi}$. Each Bell projection acts as $\proj{B_0} \otimes \identity$ on the state $U^\dagger_i\ket{\psi_{in}}\otimes \ket{\Psi}$. This leads to a required correction operation $V_i$. Telemapping requires that these $V_i$ are a tensor product of local unitary rotations. In order to test how useful the telemapping state is, one calculates the fidelity of the projection of the Bell operators $\proj{B_{i}} \otimes \identity$ onto $\ket{\psi_{in}}\ket{\Psi}$, i.e.~how well the target property $\Lambda_{out}(\psi_{in})$ is reproduced,
\begin{widetext}
$$
F=\int\sum_{i=0}^{d^2-1}\Tr\left(\proj{\psi_{in}}\otimes\left(V_i\proj{\Psi}V_i^\dagger\right)\cdot\left(\proj{B_i}\otimes \Lambda_{out}(\psi_{in})\right)\right)d\psi_{in}
$$
\end{widetext}
Replacing the $\ket{B_i}$, and using the fact that
$$
\Tr_0\left(\proj{\psi_{in}}_0\proj{B_0}_{0,1}\right)=\frac{1}{d}\proj{\psi_{in}}_1^T,
$$
leaves
$$
F=\frac{1}{d}\sum_{i=0}^{d^2-1}\Tr\left((U_i^T\otimes V_i^\dagger)R(U_i^*\otimes V_i)\proj{\Psi}\right).
$$
Hence, if the condition
\begin{equation}
[U_i^*\otimes V_i,R]\ket{\Psi}=0  \label{eqn:condition}
\end{equation}
is satisfied for all $i$, one finds that $F=d\Tr(R\proj{\Psi}),$ and the choice of $\ket{\Psi}$ that maximizes the fidelity is the maximum eigenvector of $R$, giving $F=d\lambda$, which we already know is the maximum achievable, so must be optimal, i.e.~Eqn.~(\ref{eqn:condition}) represents a sufficient condition that the optimal CP map can be implemented through telemapping.  Evidently, catalysis \cite{catalyst} (the use of a resource state to enhance the fidelity of the map) is not a meaningful concept for telemapping; such a catalyst would appear in the definition of $\ket{\Psi}$. In general, the choice of correction operations $V_i$ is not obvious {\em a priori}. In $1\rightarrow N$ cloning, the choice is clear because any rotation on the input appears on the output, and is readily canceled using $V_i=U_i^{\otimes N}$. 
However, in $2\rightarrow N$ cloning, such arguments only work in the cases where the two inputs are effectively rotated by the same unitary. Otherwise, the distribution of where each input copy appears on the output is unclear, and so we do not know how to compensate for each until the optimal map has been calculated.

{\em $1\rightarrow N$ Universal Telecloning:} When performing $1\rightarrow N$ cloning, the aim is to transform $\proj{\psi_{in}}$ to $\proj{\psi_{in}}^{\otimes N}$ with as high a fidelity as possible. Typically, two different figures of merit are applied, which manifest themselves in $\Lambda_{out}(\psi_{in})$. The first option is the global fidelity, $\Lambda_{out}(\psi_{in})=\proj{\psi_{in}}^{\otimes N}$. Instead, we shall henceforth consider the single-copy fidelity i.e.~each individual output should have as large an overlap with the input state as possible, which can be assessed using
$$
\Lambda_{out}(\psi_{in})=\sum_{n=1}^N\alpha_n\identity_{1\ldots n-1}\otimes\proj{\psi_{in}}_n\otimes\identity_{n+1\ldots N},
$$
where the $\alpha_n$ can be used to parameterize the desired asymmetry between the clones ($\sum\alpha_n=1$, $\alpha_n>0$), the fidelities $F_n$ of the clones being related by $\sum_n\alpha_nF_n=F$ \cite{review}. For universal cloning, where no prior information about the input state is available, the distribution must be taken to be uniform, and $\ket{\psi_{in}}$ can be written as $U\ket{0}$,
$$
R=\int dU\sum_{n=1}^N\alpha_nU^T\otimes U\proj{00}_{0,n}U^*\otimes U^\dagger,
$$
where the integration results in twirling \cite{horo_twirl} to give
\begin{equation}
R=\frac{1}{d(d+1)}\sum_{n=1}^N\alpha_n\left(\identity+d\proj{B_0}\right)_{0,n}.
\label{eqn:master}
\end{equation}
Eqn.~(\ref{eqn:condition}) is automatically satisfied.
For any given set of coefficients $\{\alpha_n\}$, this matrix can, in principle, be diagonalized. Given that the singlet fraction $p_{0,n}$ is intrinsically linked with the teleportation fidelity, $F_n=(p_{0,n}d+1)/(d+1)$ \cite{horo_teleport}, this elucidates the optimal trade off between how much of a singlet a particular spin can share with all the others \footnote{In this special class of cloners, our method can be understood as wanting to maximize $F=\sum_n\alpha_nF_n=\sum_n\alpha_n(p_{0,n}d+1)/(d+1)$ under the constraint $\sum_n\alpha_n=1$, which is equivalent to demanding the state $\ket{\Psi}$ which optimizes its overlap with $\sum_n\alpha_n(\proj{B_0}_{0,n}d+1)/(d+1)$. That state must be the maximum eigenvector of $R$.}.

\begin{figure}[!t]
\begin{center}
\includegraphics[width=0.4\textwidth]{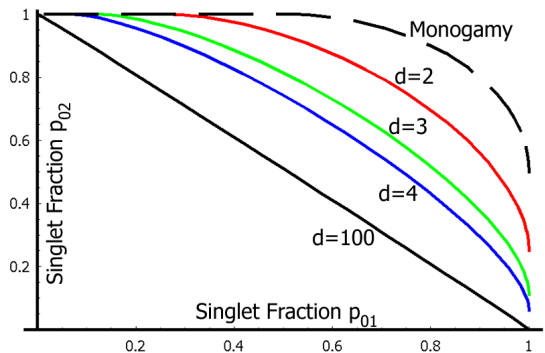}
\end{center}
\vspace{-0.5cm}
\caption{For a 3-qudit state which is maximally entangled between spin 0 and spins 1 and 2, the optimal trade-off between singlet fractions is plotted, as specified by monogamy (dashed line, qubits only) and singlet monogamy ($d=2,3,4,100$), derived from asymmetric cloning.}
\label{fig:mon_relat}
\vspace{-0.5cm}
\end{figure}

To proceed with solving Eqn.~(\ref{eqn:master}), we propose an ansatz for the maximum eigenvector of $R$,
\begin{equation}
\ket{\Psi}=\sum_{n=1}^N\beta_n\ket{B_0}_{0,n}\ket{\Phi}_{1\ldots N\neq n},
\label{eqn:guess}
\end{equation}
subject to the normalization condition
\begin{equation}
\left(\sum_{n=1}^N\beta_n\right)^2+(d-1)\sum_{n=1}^N\beta_n^2=d.
\label{eqn:norm}
\end{equation}
The state $\ket{\Phi}$ is the (normalized) uniform superposition over all permutations of $\ket{B_0}^{\otimes (N-1)/2}$ for odd $N$, and $\ket{B_0}^{\otimes (N-2)/2}\ket{0}$ for even $N$. Each covering satisfies
$$
(\proj{B_0}_{0,m}\otimes\identity)\ket{B_0}_{0,n}\ket{\Phi}=\gamma_{n,m}\ket{B_0}_{0,m}\ket{\Phi},
$$
where $\gamma_{n,m}=\left(1+\delta_{n,m}\left(d-1\right)\right)/d$, which means that $\ket{\Psi}$ is an eigenstate of $R$ provided
$$
\alpha_nd\sum_{m=1}^N\gamma_{n,m}\beta_m=(d(d+1)\lambda-1)\beta_n\qquad \forall n.
$$
Thus, to relate the $\{\alpha_n\}$ to the $\{\beta_n\}$, one just has to find the maximum eigenvector of an $N\times N$ matrix $\sum_{n,m}\alpha_n\gamma_{n,m}\ket{n}\bra{m}$. This does not prove that it is the {\em maximum} eigenvector of $R$. Let us, however, proceed under that assumption. The singlet fractions of $\ket{\Psi}$ are
$$
p_{0,n}=\left(\sum_{m=1}^N\gamma_{n,m}\beta_m\right)^2.
$$
After some rearrangement, the $\{\beta_n\}$ can be eliminated by substituting for $\{p_{0,n}\}$ in Eq.~(\ref{eqn:norm}), yielding the equality of the `singlet monogamy' relation for the singlet fractions of the cloners,
\begin{equation}
\sum_{n=1}^Np_{0,n}\leq\frac{d-1}{d}+\frac{1}{N+d-1}\left(\sum_{n=1}^N\sqrt{p_{0,n}}\right)^2.
\label{eqn:new_monogamy}
\end{equation}
The inequality can be derived by assuming equality and solving the quadratic equation for the maximum attainable value of the largest singlet fraction given the other values of singlet fraction \footnote{$\alpha_n\geq0 \Rightarrow (N+d-1)\sqrt{p_{0,n}}\geq\sum_{m=1}^N\sqrt{p_{0,m}}$, which can only be violated if $p_{0,n}\leq\frac{N-1}{d(N+d-2)}\leq 1/d$, meaning the state is PPT.}. The special case of $1\rightarrow 2$ cloning is depicted in Fig.~\ref{fig:mon_relat}.

We are now in a position to compare Eqn.~(\ref{eqn:new_monogamy}) to previous results. Setting all the $p_{0,n}$ equal returns the known result for universal symmetric cloning of
$$
F=\frac{1}{N}+\frac{2(N-1)}{N(d+1)}.
$$
Similarly, the $1\rightarrow 1+1+1$ and $1\rightarrow 1+N$ qubit cloners \cite{iblisdir-2005-72} can be found. The latter case was parametrized as
$
F_1=1-2y^2/3, F_N=\half+\frac{1}{3N}(y^2+\sqrt{N(N+2)}xy),
$
where $x^2+y^2=1$. Our solution is consistent with this, where $y^2=N(N+2)\beta_N^2/4$ and $x=\beta_1+\half N\beta_N$. Thus, we know that at least at certain points of the phase diagram, $\ket{\Psi}$ is the maximum eigenvector. This has also been confirmed analytically for $d=2, N\leq5$ and $d\leq 5, N=3$ for all asymmetries.

{\em Application of Singlet Monogamy:} There are many situations where singlet fraction is a more relevant parameter to estimate than the tangle, and thus no-cloning bounds give much tighter results. Consider, for example, the Heisenberg model on a regular lattice of coordination number $c$ and N spins.
$$
H_{\text{Heis}}=\smallfrac{4}\sum_{\langle i,j\rangle}(XX+YY+ZZ)_{i,j},
$$ for which we might like to bound the ground state energy. The ground state can be taken to be $\ket{\Psi}$, with energy per site $E=\bra{\Psi}H_{\text{Heis}}\ket{\Psi}/N$. However, this can be rephrased as simply the sum of singlet fractions of $\ket{\Psi}$ for all nearest-neighbor pairs,
$$
EN=\smallfrac{8}Nc-\sum_{\langle i,j\rangle}p_{i,j}.
$$
The ground state must reproduce the translational invariance of the Hamiltonian, so all the singlet fractions are equal, $E=\half c(1/4-p)$. By assuming monogamy, the tangle possible between a pair of neighboring sites is $\tau\leq 1/c$, which yields $p\leq\half(1+1/\sqrt{c})$. By contrast, the singlet monogamy relation for qubits reveals that $p\leq\half(1+1/c)$, giving a much tighter bound for E. This same bound has previously been achieved in \cite{alternate}, which used the technique of dividing the lattice into small repeating units, and diagonalizing the corresponding Hamiltonian \cite{anderson} -- the sum of ground state energies of blocks of terms is a lower bound to the overall ground state energy. These small blocks of Hamiltonian are precisely those used for the calculation of the optimal cloner in Eqn.~(\ref{eqn:master}), so the apparent coincidence is no such thing. Differing coupling strengths along different spatial directions can be accounted for using asymmetric cloning, and performing an optimization over the asymmetry parameters.

A feature of our formulation is that spin 0 is taken to be maximally entangled with the other spins. In comparison, the monogamy relation of Eqn.~(\ref{eqn:tangle}) allows an arbitrary value for $\tau(\rho_{0,1\ldots N})$, although it is often hard to determine, and commonly set to its maximal value of 1 for qubits. For a translationally invariant spin-$\half$ system with magnetization $\langle S_{\vec n}\rangle$ along direction $\vec n$, the tangle $\tau(\rho_{0,1\ldots c})\leq(1-\langle S_{\vec n}\rangle^2)^2$, which can be used to impose a bound on the singlet fraction, and thus the validity of a mean-field approximation of the energy of a Heisenberg Hamiltonian,
$$
\varepsilon\leq p_{0,1}-\half\leq(1-\langle S_{\vec n}\rangle^2)/\sqrt{4c}.
$$
Crucially, as the coordination number increases, the accuracy, $\varepsilon$, of the mean-field approximation improves. Although singlet monogamy has no way to incorporate the bound on $\tau$, we still arrive at
$$
\varepsilon\leq p_{0,1}-\half\leq1/(2c),
$$
which is a better bound for $\langle S_{\vec n}\rangle^2\leq 1-1/\sqrt{c}$, proving that the mean-field approximation converges even faster with increasing coordination number. Potentially, one could choose a telecloning state $\ket{\Psi}$ such that $\Tr(R\proj{\Psi})$ is maximum under the constraint that $\ket{\Psi}$ has some specific entanglement, which would serve to relax this property. This is left open for future study.

{\em Conclusions:} In this paper, we have given a sufficient condition that the optimal implementation of a CP map can be achieved through telemapping. The problem of universal $1\rightarrow N$ cloning with arbitrary asymmetries and spin dimensions is one such example, and thus the calculation of the optimal fidelity is reduced to the diagonalization of a simple matrix. Strong analytical and numerical evidence suggests that Eq.~(\ref{eqn:guess}) gives the maximum eigenvector of this matrix, yielding Eq.~(\ref{eqn:new_monogamy}), a new monogamy relation for singlet fractions for qudits of any local Hilbert space dimension. Singlet fraction is a physically relevant parameter in many settings, and such a relation can significantly outperform the original monogamy inequality, which is only applicable to qubits. We have demonstrated this with some simple examples for calculating the properties of Heisenberg Hamiltonians. 

We note that this formalism is not limited to solving the case of universal cloning. Instead, one can introduce an arbitrary distribution function. In the case of qubits, with $\ket{\psi_{in}}=\cos(\theta/2)\ket{0}+\sin(\theta/2)e^{i\phi}\ket{1}$, restricting to the distribution function $f(\theta,\phi)=f(\theta)$ with $\int_0^{\pi}\cos\theta f(\theta)d\theta=0$ is sufficient to satisfy Eqn.~(\ref{eqn:condition}), and it turns out that a single state suffices to clone all such cases where the classical bound can be exceeded. Furthermore, the state can be efficiently produced on a quantum computer \cite{kaye}. The formalism also does not allow for any ancilla qubits, which means that no `anti clones' appear, as has been the case in previous studies. The interesting questions for the future are whether the formalism can be applied to other CP maps that one may want to implement, what monogamy equivalents these yield, and whether the singlet monogamy relation can be adapted for the non-maximally entangled situation.

{\em Acknowledgments:} This work was supported by the National Research Foundation \& Ministry of Education, Singapore. AK is also supported by DFG Cluster of Excellence Munich-Centre for Advanced Photonics (MAP) and Clare College, Cambridge.

\end{document}